\documentclass{ifacconf}

\usepackage{graphicx}      
\usepackage{natbib}        
\usepackage{amsfonts}
\usepackage{amsmath,amssymb}
\usepackage{color}
\usepackage{mathrsfs} 

\newcommand{\vel}{\mathbf{v}} 
\newcommand{\den}{\rho} 
\newcommand{\ten}{\boldsymbol{\tau}} 
\newcommand{\hf}{\mathbf{q}} 
\newcommand{\visc}{\mu}
\newcommand{\visd}{\kappa}
\newcommand{\s}{s} 
\newcommand{\p}{p} 
\newcommand{\vor}{\boldsymbol{\omega}}
\newcommand{\vorD}{\omega}
\newcommand{\V}{\Omega} 
\newcommand{\Vb}{\partial\Omega}
\newcommand{\h}{h}

\newcommand{\tena}{\boldsymbol{\sigma}} 	

\newcommand{\Ha}{\mathcal{H}}

\newcommand{\er}{e_{\den}}
\newcommand{\ev}{\mathbf{e}_{\vel}}
\newcommand{\es}{e_{\s}}

\newcommand{\J}{\mathcal{J}}
\newcommand{\Jt}{\mathcal{J}_{\boldsymbol{\tau}}}
\newcommand{\Jq}{\mathcal{J}_{\hf}}
\newcommand{\Gt}{\mathcal{G}_{\boldsymbol{\tau}}}
\newcommand{\Gw}{\mathcal{G}_{r}}
\newcommand{\Gr}{\mathcal{G}_{d}}
\newcommand{\eb}{\mathbf{e}_{\partial}}
\newcommand{\fb}{\mathbf{f}_{\partial}}
\newcommand{\ew}{\mathbf{e}_{r}}
\newcommand{\fw}{\mathbf{f}_{r}}
\newcommand{\ed}{e_{d}}
\newcommand{\fd}{f_{d}}
\newcommand{\n}{\mathbf{n}}
\newcommand{\un}{\mathbf{u}}

\newcommand{\grad}{\mathbf{grad}~} 
\renewcommand{\div}{{\rm div}~}	
\newcommand{\curl}{\mathbf{curl}~} 
\newcommand{\Grad}{\mathbf{Grad}~} 
\newcommand{ \Div}{\mathbf{Div}~}	
\newcommand{\dt}[1]{\partial_{t}#1}
\newcommand{\inner}[2]{\left\langle #1 \right\rangle_{#2}}
\newcommand{ \T}[0]{{\rm T}}

\theoremstyle{plain}
\newtheorem{definition}{Definition}
\newtheorem{proposition}{Proposition}
\newtheorem{rmk}{Remark}
\begin{document}
\begin{frontmatter}

\title{About dissipative and pseudo Port-Hamiltonian Formulations of irreversible Newtonian Compressible Flows} 

\thanks[footnoteinfo]{This project has received funding from the European Union's Horizon 2020 research and innovation programme under the Marie Sklodowska-Curie fellowship, ConFlex ITN Network	and by the INFIDHEM project under the reference codes 765579 and ANR-16-CE92-0028 respectively, also by	CONICYT through grands CONICYT-PFCHA/Bec. Doc. Nac./2017-21170472, FONDECYT 1181090, FONDECYT 1191544 and BASAL FB0008.}

\author[First,Third]{Luis A. Mora} 
\author[First]{Yann Le Gorrec} 
\author[Second]{Denis Matignon}
\author[Third]{Hector Ramirez}
\author[Third]{Juan I. Yuz}

\address[First]{FEMTO-ST Institute, AS2M department ,Univ. Bourgogne Franche-Comt\'{e}, Univ. de Franche-Comt\'{e}/ENSMM, 24 rue Savary, F-25000 Besan\c{c}on, France.}
\address[Second]{ISAE-SUPAERO,Universit\'{e} de Toulouse, 10 Avenue Edouard Belin, BP-54032, Cedex 4, Toulouse 31055, France.}
\address[Third]{AC3E, Universidad T\'{e}cnica Federico Santa Mar\'{i}a, Av. Espa\~{n}a 1680, Valparaiso, Chile.}

\begin{abstract}                
In this paper we consider the problem of obtaining a general port-Hamiltonian formulation of Newtonian fluids. We propose the port-Hamiltonian models to describe the energy flux of rotational three-dimensional isentropic and non-isentropic fluids, whose boundary flows and efforts can be used for control purposes or for power-preserving interconnection with other physical systems. In case of two-dimensional flows, we include the considerations about the operators associated with fluid vorticity, preserving the port-Hamiltonian structure of the models proposed.
\end{abstract}


\end{frontmatter}

\section{Introduction}
In control theory, models are required  to describe the plant dynamics with sufficient precision and simplicity. In particular, energy-based control methods, such as energy-shaping \citep{Macchelli2017}, IDA-PBC \citep{Vu2015}, observer-based control \citep{Toledo2019}, among others, require models describing the energy flux of the physical phenomena. These models are commonly formulated using the port-Hamiltonian (PH)  framework.

Port-Hamiltonian  systems provides useful properties for the control theory, such as passivity, stability in the Lyapunov sense and power-preserving connectivity by ports \citep{VanderSchaft2014}. For infinite-dimensional systems a PH formulation based in a Stokes-Dirac structures is proposed by \cite{LeGorrec2005} and an extension to include dissipative effects is presented in \cite{Villegas2006}. As soon as irreversible thermodynamics systems are considered, the PH formulation are not valid anymore. In the finite-dimensional case a first intend to cope in a structural way with this class of systems is the irreversible port-Hamiltonian formulation \citep{Ramirez2013}, but alternative approaches using pseudo-PH system have also been proposed. 

In the current paper we focus in the dynamics and thermodynamics of non-reactive Newtonian fluids. This kind of fluids has been studied in different areas of engineering applications, from biomedical systems, as the phono-respiratory modeling \citep{Mora2018}, to Fluid-Structure-Interaction problems \citep{Cardoso-Ribeiro2017}.

Different energy-based approaches  have been presented in literature to describe Newtonian fluids. However, these approaches are constrained to a kind of fluid due to the assumptions that were considered. For example, for ideal isentropic fluids, 1D PH models are proposed by \cite{Macchelli2017} for inviscid fluids and \cite{Kotyczka2013} with friction dissipation, for control purposes and  pipe network modeling, respectively, where the voricity effects are neglected as a consequence of the one-dimensional assumption. In this sense, a Hamiltonian model based on stream functions to describe the vorticity dynamics of 2D fluid is presented in \cite{Swaters2000}, however the model is limited to potential flows. A dissipative PH model of 3D irrotational fluids is proposed by \cite{Matignon2013} and a general Hamiltonian model for inviscid fluid is presented by \cite{VanderSchaft2002}. For non-isentropic fluids a one-dimensional model for reactive flows is proposed by \cite{Altmann2017}, neglecting the vorticity effects.


In this work we present a general energy-based formulation for  isentropic and non-isentropic 3D compressible fluids using a pseudo-PH framework, including the vorticity effects in the velocity field. First we develop a pseudo-PH model for non-isentropic fluids, focusing on non-reactive flows. Later, we describe the treatment of terms associated with the viscous tensor under an isentropic assumption for the fluid, to obtain a dissipative PH model. Finally, we describe the necessary considerations to conserve the PH structure of the models proposed for 2D fluids. 

The current paper is organized as follows: In Section \ref{sec:Non-Isen} an infinite-dimensional pseudo PH model is developed for non-isentropic fluids. In Section \ref{sec:Isen}, we consider an isentropic assumption, rewriting the term associated with the viscous stress tensor to obtain a dissipative PH model. Section \ref{sec:2D} describes the considerations over the operators associated with the vorticity, to conserve the same structure of PH models developed in previous sections. Finally, the conclusions are presented in Section \ref{sec:Conclu}. The notation and mathematical identities are summarized in the Appendix.

\section{Non-isentropic fluid}\label{sec:Non-Isen}
In this section we describe the energy-based formulation for non-insentropic fluids.
Denote by $\den$, $\vel$, $\s$ and $ T$ the density, velocity field, entropy per unit of mass, and temperature of the fluid, respectively. The fluid dynamics are described by the following governing equations:
\begin{subequations}
	\begin{align}
		\dt{\den}&=-\div{\den\vel}\label{eq:Con1}\\
		\den\dt{\vel}&=-\den\vel\cdot\Grad{\vel}-\grad{\p}- \Div{\ten}\label{eq:Mot1}\\
		\den T\dt{\s}&=-\den T\vel\cdot\grad{s}-\ten:\Grad{\vel}-\div{\hf}\label{eq:Heat1}
	\end{align}
\end{subequations}
where \eqref{eq:Con1} and \eqref{eq:Mot1} are the continuity and motion equations, respectively, and \eqref{eq:Heat1} is the general equation of heat transfer \citep{Landau1987}; $p$ is the static pressure, $\ten$ is the viscosity tensor and  $\hf$ is the heat flux. In this work, we consider non-reactive Newtonian fluids. Then, $\ten$ and $\hf$ are defined as:
\begin{align}
\ten=&-\visc\left(\Grad{\vel}+\left[\Grad{\vel}\right]^{ \T}-\frac{2}{3}\left(\div{\vel}\right)I\right)\nonumber\\ 	
	 &-\visd\left(\div{\vel}\right)I\label{eq:Ntensor}\\
\hf=&-K\grad{ T}\label{eq:hf}
\end{align}
where $\visc$ and $\visd$ are the shear and dilatational viscosities \citep{Bird2015}, respectively, $I$ is the identity matrix and $K$ is a non-negative matrix that describes the thermal conductivity of the fluid \citep{Ottinger2005}. 

In fluid dynamics the tendency to rotate is characterized by the vorticity $\vor=\curl{\vel}$ and the term $\vor\times\vel$, from the point of view of energy,  describes the power exchange between the velocity field components given by the fluid rotation.

\begin{definition}
	Let $\vor=\begin{bmatrix}
	\vorD_1 & \vorD_2 & \vorD_3
	\end{bmatrix}^{ \T}$ the vorticity vector of the fluid. We define the fluid  {Gyroscope} $G_{\vor}$ as the skew-symmetric matrix, such that $G_{\vor}\vel=\vor\times\vel$. For 3D fluids, the  {Gyroscope} is given by:
	\begin{align}
	G_{\vor}&=\begin{bmatrix}
	0 & -\vorD_3 & \vorD_2\\
	\vorD_3 & 0 & -\vorD_1\\
	-\vorD_2 & \vorD_1 & 0
	\end{bmatrix}
	\end{align}
	 
\end{definition}

On the other hand in\eqref{eq:Heat1}, the term $ \div{\hf}$ can be rewritten as 
\begin{align}
	 \div{\hf}=T \div{\hf_s}+\hf_s\cdot\grad{T}\label{eq:hf2}
\end{align}
 where $\hf_s$ is the entropy flux by heat conduction \citep{Bird2015}. For non-reactive fluids $\hf_s$ is defined by
 \begin{align}
 \hf_s&=-\frac{K}{T}\grad{T}\label{eq:hfs}
 \end{align}
 Then, considering the Gibbs equation 
\begin{align}
du=-pd\left(\frac{1}{\den}\right)+ T d\s
\end{align}
that describes the change of the specific internal energy $u$ with respect to changes of $\den$ and $\s$, the fluid enthalpy $h=u+p/\den$ and the relationships $\grad{\frac{1}{\den}p}=\frac{1}{\den}\grad{p}+p\grad{\frac{1}{\den}}$ and $T=\partial_{\s}u$. Then, we can rewrite the fluid dynamics in terms of the state variables and the temperature, namely:
\begin{subequations}
	\begin{align}
	\dt{\den}=&-\div{\den\vel}\\
	\dt{\vel}=&-\grad{\left(\frac{1}{2}\vel\cdot\vel+\h\right)}-G_{\vor}\vel+T\grad{s}\nonumber\\
	&-\frac{1}{\den} \Div{\ten}\\
	\dt{\s}=&-\vel\cdot\grad{\s}-\frac{\ten}{\den T}:\Grad{\vel}-\frac{\hf_s}{\den T}\cdot\grad{T}\nonumber\\
			&-\frac{1}{\den}\div{\hf_s}\label{eq:entropybalance1}
	\end{align}\label{eq:fluid}
\end{subequations}

Note that entropy generation, second law of thermodynamics, is given by the following non-negative condition \citep{Ottinger2005}:
\begin{align}
	-\frac{1}{\den T}\ten:\Grad{\vel}-\frac{\hf_s}{\den T}\cdot\grad{T}\geq 0
\end{align}
where $-\frac{1}{\den T}\ten:\Grad{\vel}$ is the rate of entropy creation by the kinetic energy dissipated into heat by viscosity friction, and $-\frac{\hf_s}{\den T}\cdot\grad{T}$ is the rate of entropy creation by heat flux.

\subsection{Port-Hamiltonian description of non-isentropic fluids}

Consider the fluid domain $\V$ with boundary $\Vb$.
The total energy of the fluid described in \eqref{eq:fluid} is given by:
\begin{align}
	\Ha&=\int_{\V}\frac{1}{2}\den\vel\cdot\vel+\den u\left(\den,\s\right)\label{eq:HNi}
\end{align}

Then, the fluid efforts $\mathbf{e}=\begin{bmatrix}
\er & \ev^{ \T} & \es
\end{bmatrix}^{ \T}$ are given by the variational derivative of the energy, namely
\begin{align}
	\begin{bmatrix}
	\er\\	\ev\\	\es
	\end{bmatrix}&=\begin{bmatrix}
	\delta_{\den}\Ha\\ \delta_{\vel}\Ha\\\delta_{\s}\Ha
	\end{bmatrix}=\begin{bmatrix}
		\frac{1}{2}\vel\cdot\vel+h\\
		\den\vel\\
		\den T
	\end{bmatrix}\label{eq:efforts}
\end{align}

Note that $\delta_{\den}\Ha=\frac{1}{2}\vel\cdot\vel+u+\den\partial_{\den}u$. Given the relationship $p=\den^2\partial_{\den}u$, we obtain  $\delta_{\den}\Ha=\frac{1}{2}\vel\cdot\vel+u+p/\den=\frac{1}{2}\vel\cdot\vel+h$. Using \eqref{eq:efforts}, the fluid dynamics in \eqref{eq:fluid} can be related with energy through the fluid efforts, i.e.,
\begin{subequations}
	\begin{align}
		\dt{\den}=&-\div{\ev}\\
		\dt{\vel}=&-\grad{\er}-\frac{G_{\vor}}{\den}\ev+\grad{s}\frac{\es}{\den}\nonumber\\
		&-\frac{1}{\den} \Div{\frac{\ten}{\den T}\es}
	\end{align}
	\begin{align}
	\dt{\s}=&-\grad{\s}\cdot\frac{\ev}{\den}-\frac{1}{\den T}\ten:\Grad{\frac{\ev}{\den}} \label{eq:entropybalance2}\\
			&+\frac{1}{\den T}\left\|\grad{\frac{\es}{\den}}\right\|^2_{\frac{K}{T}} +\frac{1}{\den}\div{\left[\frac{K}{T}\grad{\frac{\es}{\den}}\right]}\nonumber
	\end{align}\label{eq:fluid2}
\end{subequations}

To obtain the port-Hamiltonian formulation, it is necessary set the interconnections between the components of the fluid dynamics. In the case of the velocity field and the entropy, they are interconnected through the operator $\Jt$ and the corresponding adjoint $\Jt^*$ in the effort space. Similarly, the last two terms of the right-hand side in \eqref{eq:entropybalance2} describes the rate of irreversible entropy creation and entropy diffusion by the heat flux $\hf$, thus, these phenomena can be characterized using an operator $\Jq$. The above operators are defined in the following Lemmas.

\begin{lem}\label{lem:1}
Let $\ten$ be a symmetric second order tensor and $\Jt=\grad{s}\left(\frac{\cdot}{\den}\right)-\frac{1}{\den} \Div{\left(\frac{\ten}{\den T}\cdot\right)}$ an operator on the entropy effort $\es$. Then, the adjoint operator $\Jt^*$ in the effort space of the fluid is given by $\Jt^*=\grad{s}\cdot\left(\frac{\cdot}{\den}\right)+\frac{\ten}{\den T}:\Grad{\frac{\cdot}{\den}}$, such that
\begin{align}
\inner{\ev,\Jt\es}{\V}-\inner{\es,\Jt^*\ev}{\V}&=-\int_{\Vb}\ten:\left[\frac{\ev}{\den}\n^{ \T}\right]\label{eq:b1}
\end{align}
\end{lem}
\begin{pf}
	Consider the inner product
	\begin{align}
	\inner{\ev,\Jt\es}{\V}=&\int_{\V}\ev\cdot\Jt\es \nonumber\\
	=&\int_{\V} \ev\cdot\left[\grad{s}\frac{\es}{\den}-\frac{1}{\den} \Div{\left[\frac{\ten}{\den T}\es\right]}\right] \nonumber
	\end{align}
	Then, using the identity \eqref{eq:B2}, where $\tena={\ten\es}/{\den T}$ and $\un={\ev}/{\den}$, the inner product in previous equation can be rewritten as:
	\begin{align}
	\inner{\ev,\Jt\es}{\V}=&\int_{\V}\left[\grad{s}\cdot\frac{\ev}{\den}+\frac{\ten}{\den T}:\Grad{\frac{\ev}{\den}}\right]\es  \nonumber\\
	&-\int_{\Vb}\left[ \frac{\ten \es}{\den T}\cdot \frac{\ev}{\den}\right]\cdot\n \nonumber\\
	=&\inner{\Jt^*\ev,\es}{\V}-\int_{\Vb}\left[ \frac{\ten \es}{\den T}\cdot \frac{\ev}{\den}\right]\cdot\n\label{eq:pf1}
	\end{align}
	where $\Jt^*=\grad{s}\cdot\left(\frac{\cdot}{\den}\right)+\frac{\ten}{\den T}:\Grad{\frac{\cdot}{\den}}$ and $\n$ is the normal outward unitary vector to the boundary $\Vb$. Considering the boundary conditions equal to 0 we obtain that $\inner{\ev,\Jt\es}{\V}=\inner{\Jt^*\ev,\es}{\V}$, i.e., $\Jt^*$ is the formal adjoint of $\Jt$.	
	Finally, using the mathematical identity $\left(\ten\cdot\vel\right)\cdot\n=\ten:\vel\n^{ \T}$ we obtain
	\begin{align}
	\inner{\ev,\Jt\es}{\V}-\inner{\es,\Jt^*\ev}{\V}=&-\int_{\Vb}\ten:\left[\frac{\ev}{\den}\n^{ \T}\right]
	\end{align}
	where $\frac{\ev}{\den}\n^{ \T}$ is the tangential projection of the velocity field.	\hfill\hfill \qed 
\end{pf}

\begin{lem}\label{lem:2}
	Let $\Jq$ be an operator on space of the entropy effort $\es$, defined as
	\begin{align}
	\Jq=\mathcal{Q}_T-\mathcal{G}^*_TS_{T}\mathcal{G}_T,
	\end{align}
	where $Q_T$ and $\mathcal{G}^*_TS_{T}\mathcal{G}_T$ describes two phenomena associated with the heat flux. $\mathcal{Q}_T=\frac{1}{\den T}\left\|\grad\frac{\cdot}{\den}\right\|^2_{S_T}$  describes the entropy creation, such that $\mathcal{Q}_T\es\geq0,\forall\es$, and $-\mathcal{G}^*_TS_T\mathcal{G}_T$ describes the entropy diffusion, where the operator $\mathcal{G}^*_T=\frac{1}{\den}\div$ is the formal adjoint of $\mathcal{G}_T=-\grad{\frac{\cdot}{\den}}$ and $S_T=\frac{K}{T}\geq0$. Then, the rate of entropy addition by heat flux can be expressed as
	\begin{align}
	-\frac{1}{\den T}\div{\hf}=&\Jq\es
	\end{align}
	satisfying
	\begin{align}
	\inner{\es,\Jq\es}{\V}&=-\int_{\Vb}T\left(\hf_s\cdot\n\right)\label{eq:b2}
	\end{align} 
\end{lem}
\begin{pf}
	Note that $\frac{1}{\den T}\div{\hf}=\frac{\hf_s}{\den T}\cdot\grad{T}+\frac{1}{\den}\div{\hf_s}$. Defining $S_T=K/T$, from \eqref{eq:entropybalance2} we obtain
	\begin{align}
		-\frac{\hf_s}{\den T}\cdot\grad{T}&
				=\frac{1}{\den T}\left\|\grad{\frac{\es}{\den}}\right\|^2_{S_T}\label{eq:pf21}\\
		-\frac{1}{\den}\div{\hf_s}&=\frac{1}{\den}\div{\left[S_T\grad{\frac{\es}{\den}}\right]} \label{eq:pf2}
	\end{align}
	Given that the formal adjoint of divergence is minus the gradient, it is easy to prove that $\mathcal{G}^*_T=\frac{1}{\den}\div{}$ is the formal adjoint of operator $\mathcal{G}_T=-\grad{\frac{\cdot}{\den}}$. Then, from \eqref{eq:pf21} and \eqref{eq:pf2} the entropy addition by heat flux can be expressed as
	\begin{align}
	-\frac{1}{\den T}\div{\hf}
	=&\left(\mathcal{Q}_T-\mathcal{G}^*_TS_T\mathcal{G}_T\right)\es=\Jq\es\label{eq:pf3}
	\end{align}
	The inner product in the left-hand side of \eqref{eq:b2}, is given by
	\begin{align}
	\inner{\es,\Jq\es}{\V}=&\int_{\V}\frac{\es}{\den T}\left\|\grad{\frac{\es}{\den}}\right\|^2_{S_T}\nonumber\\
	&+\int_{\V}\frac{\es}{\den}\div{\left[S_T\grad{\frac{\es}{\den}}\right]}\nonumber\\
	=&-\int_{\V}\left[\grad \frac{\es}{\den}\right]\cdot\left[-S_T\grad \frac{\es}{\den}\right]\nonumber\\
	&-\int_{\V}\frac{\es}{\den}\div{\left[-S_T\grad{\frac{\es}{\den}}\right]}\nonumber\\
	=&-\int_{\V}\hf_s\cdot\grad{\frac{\es}{\den}}+\frac{\es}{\den}\div{\hf_s}\nonumber
	\end{align}
	Finally, using the property \eqref{eq:B2} we obtain \eqref{eq:b2}.
	\hfill\hfill\qed 
\end{pf}

\begin{lem}
	Let $\mathcal{E}_s=\{\forall\es:\es/\den T=1\}$ be the space of entropy efforts. Then, the operator $\Jq=\mathcal{Q}_T-\mathcal{G}_T^{*}S_T\mathcal{G}_T$ is skew-symmetric in $\mathcal{E}_s$.
\end{lem}
\begin{pf}
	Let be the efforts $e_{s,1}\in\mathcal{E}_s$ and $e_{s,2}\in\mathcal{E}_s$. Then, we obtain:
	\begin{align}
	\inner{e_{s,1},\Jq e_{s,2}}{\V}=&\int_{\V}e_{s,1}\mathcal{Q}_Te_{s,2}- e_{s,1}\mathcal{G}^{*}_TS_T\mathcal{G}_T e_{s,2}\nonumber\\
		=&\int_{\V}\frac{e_{s,1}}{\den T}\left\| 	\grad{\frac{e_{s,2}}{\den}}\right\|_{S_T}^2\nonumber\\
		 &+\frac{e_{s,1}}{\den}\div{S_T\grad{\frac{e_{s,2}}{\den}}}\nonumber\\
		=&-\int_{\V}\hf^2_s\cdot\grad{\frac{e_{s,2}}{\den}}+\frac{e_{s,1}}{\den}\div{\hf_s^2}\label{eq:pf41}
	\end{align}
	where $\hf_{s}^j=-S_T\grad{\dfrac{e_{s,j}}{\den}}$, $j\in\{1,2\}$. Given that $\{e_{s,1},e_{s,2}\}\in\mathcal{E}_s$, the term $\grad{\dfrac{e_{s,2}}{\den}}$ can be rewritten as
	\begin{align}
	\grad{\frac{e_{s,2}}{\den}}=&\grad{\frac{e_{s,1}e_{s,2}}{\den^2 T}}\nonumber\\
	=&\frac{e_{s,2}}{\den T}\grad{\frac{e_{s,1}}{\den}}+\frac{e_{s,1}}{\den}\grad{\frac{e_{s,2}}{\den T}}\nonumber\\
	=&\grad{\frac{e_{s,1}}{\den}}\label{eq:pf42}
	\end{align}
	Substituting \eqref{eq:pf42} in \eqref{eq:pf41} we obtain
	\begin{align}
	\inner{e_{s,1},\Jq e_{s,2}}{\V}=&-\int_{\V}\hf^2_s\cdot\grad{\frac{e_{s,1}}{\den}}+\frac{e_{s,1}}{\den}\div{\hf_s^2}\nonumber\\
	\overset{\eqref{eq:A3}}{=}&-\int_{\V}\div{\left[\frac{e_{s,1}}{\den}\hf_s^2\right]}\label{eq:pf43}
	\end{align}
	Similarly the inner product $\inner{\Jq e_{s,1},e_{s,2}}{\V}$ is given by
	\begin{align}
	\inner{\Jq e_{s,1},e_{s,2}}{\V}=&-\int_{\V}\div{\left[\frac{e_{s,2}}{\den}\hf_s^1\right]}\label{eq:pf44}
	\end{align}
	Now, rewritten the term $\frac{e_{s,1}}{\den}\hf_s^{2}$ in \eqref{eq:pf43} as
	\begin{align}
	\frac{e_{s,1}}{\den}\hf_s^2=&-\frac{e_{s,1}}{\den}S_T\grad{\frac{e_{s,2}}{\den}} \nonumber\\
	=&-\left[S_T\grad{\frac{e_{s,1}e_{s,2}}{\den^2}}-\frac{e_{s,2}}{\den}S_T\grad{\frac{e_{s,1}}{\den}}\right]\nonumber\\
	=&-S_T\grad{\frac{e_{s,1}e_{s,2}}{\den^2}}-\frac{e_{s,2}}{\den}\hf_s^1\label{eq:pf45}
	\end{align}
	Then, equation \eqref{eq:pf43} can be expressed as
	\begin{align}
	\inner{e_{s,1},\Jq e_{s,2}}{\V}
	=&\int_{\V}\div{\left[S_T\grad{\frac{e_{s,1}e_{s,2}}{\den^2}}+\frac{e_{s,2}}{\den}\hf_s^1\right]}\nonumber\\
	=&\int_{\V}\div{\left[\frac{e_{s,2}}{\den}\hf_s^1\right]}\nonumber\\
	&+\int_{\V}\div{\left[S_T\grad{\frac{e_{s,1}e_{s,2}}{\den^2}}\right]}\nonumber\\
	=&-\inner{\Jq e_{s,1},e_{s,2}}{\V}\nonumber\\
	&+\int_{\Vb}\left[S_T\grad{\frac{e_{s,1}e_{s,2}}{\den^2}}\right]\cdot\n\label{eq:pf46}
	\end{align}
	Then, considering the boundary conditions equal to 0, $\inner{e_{s,1},\Jq e_{s,2}}{\V}=-\inner{\Jq e_{s,1},e_{s,2}}{\V}$. i.e., $\Jq$ is a formal skew-adjoint operator.\hfill\hfill\qed
\end{pf}

Thus, using the above Lemmas, the fluid dynamics for non-isentropic fluids can be expressed as an energy-based model, as we shown in the next proposition.

\begin{proposition}\label{prop:nonisen}
	Consider a non-isentropic Newtonian compressible fluid, whose total energy is describe by \eqref{eq:HNi}. Then, the governing equations in \eqref{eq:fluid} can be expressed as pseudo infinite-dimensional port-Hamiltonian system
	\begin{align}
	\dt{\boldsymbol{x}}&=\J	\mathbf{e}\label{eq:PHNonisen}
	\end{align}
	where $\boldsymbol{x}=\begin{bmatrix}
	\den & \vel^{ \T} &\s
	\end{bmatrix}^{ \T}$ is the state vector, $\mathbf{e}=\begin{bmatrix}
	\er & \ev^{ \T} & \es
	\end{bmatrix}^{ \T}$ is the fluid effort vector described in \eqref{eq:efforts}, $\mathbf{f}_R=\mathcal{G}_T\es$ and $\mathbf{e}_R=S_T\mathbf{f}_R$ are the flow and effort associated with entropy diffusion, and $\J$  is an operator given by
	\begin{align}
	\J&=\begin{bmatrix}
	0 & -\div & 0 \\
	-\grad &-\frac{1}{\den}G_{\vor} & \Jt\\
	0 & -\Jt^* & \mathcal{Q}_T -\mathcal{G}^{*}_TS_T\mathcal{G}_T
	\end{bmatrix}
	\end{align}
	satisfying
	\begin{align}
	\dot{\Ha}=\inner{\eb,\fb}{\Vb}\label{eq:PowBalNI}
	\end{align}
	where $\inner{\eb,\fb}{\Vb}$ is the power supplied through the boundary $\Vb$ and
	the boundary flows $\fb$ and efforts $\eb$ are given by $$\fb=\begin{bmatrix}
	-(\ev\cdot\n)|_{\Vb}\\-\vel\n^{ \T}|_{\Vb}\\-(\hf_s\cdot\n)|_{\Vb}
	\end{bmatrix} \text{ and } \eb=\begin{bmatrix}
	\er|_{\Vb}\\ \ten|_{\Vb}\\ T|_{\Vb}
	\end{bmatrix}.$$

\end{proposition}
\begin{pf}
	The fluid governing equations in \eqref{eq:fluid} can be rewritten as function of the fluid efforts described in \eqref{eq:efforts}, as shown in \eqref{eq:fluid2}. Then, using the operators defined in Lemmas \ref{lem:1} and \ref{lem:2} we obtain
	\begin{align}
		\underset{\dt{\boldsymbol{x}}}{\underbrace{\begin{bmatrix}
		\dt{\den}\\\dt{\vel}\\\dt{\s}
		\end{bmatrix}}}&=\underset{\J}{\underbrace{\begin{bmatrix}
		0 & -\div& 0\\
		-\grad &-\frac{1}{\den}G_{\vor} & \Jt\\
		0 & -\Jt^* & \mathcal{Q}_T-\mathcal{G}^*_TS_{T}\mathcal{G}_T
		\end{bmatrix}}}\underset{\mathbf{e}}{\underbrace{\begin{bmatrix}
		\er\\\ev\\\es
		\end{bmatrix}}}
	\end{align}
	The energy balance for this system is given by:
	\begin{align}
	\dot{\Ha}=&\inner{\mathbf{e},\J\mathbf{e}}{\V}=\int_{\V}\mathbf{e}\cdot\J\mathbf{e}\nonumber\\
	=&-\int_{\V}\er\div{\ev}+\ev\cdot\grad{\er}-\frac{\ev}{\den}\cdot G_{\vor}\ev \nonumber\\
		&+\inner{\ev,\Jt\es}{\V}-\inner{\es,\Jt^*\ev}{\V}+\inner{\es,\Jq\es}{\V}\label{eq:pfdH}
	\end{align}
	
	Note that given the skew-symmetry property of the gyroscope $\frac{\ev}{\den}G_{\vor}\ev=0$. Then, using \eqref{eq:b1} and \eqref{eq:b2}, equation \eqref{eq:pfdH} can be rewritten as
	\begin{align}
	\dot{\Ha}=&-\int_{\Vb}\er\left(\ev\cdot\n\right)+\ten:\left[\frac{\ev}{\den}\n^{ \T}\right]+T\left[\hf_s\cdot\n\right]\label{eq:pf51}
	\end{align}
	Defining the boundary flows and efforts as:
	\begin{align}
	\fb&=\begin{bmatrix}
	\left.-\left(\ev\cdot\n\right)\right|_{\Vb} \\ \left. -\frac{\ev}{\den}\n^{ \T}\right|_{\Vb}\\ \left.-\left(\hf_s\cdot\n\right)\right|_{\Vb}
	\end{bmatrix}& \eb&=\begin{bmatrix}
	\left.\er\right|_{\Vb}\\\ten|_{\Vb}\\\left.\frac{\es}{\den}\right|_{\Vb}
	\end{bmatrix} 
	\end{align}
	where $\ev\cdot\n$ is the normal projection of the momentum density, $\frac{\ev}{\den}\n^{ \T}$ is the tangential projection of the velocity field. Then, the rate of change of the total energy is given by:
	\begin{align}
	\dot{\Ha}=\inner{e_{\partial},f_{\partial}}{\Vb}
	\end{align}
	\hfill\hfill\qed
\end{pf}

\begin{rmk}
	Note that the system in \eqref{eq:PHNonisen} looks like a Stokes-Dirac structure because of the skew symmetry of the operators involved, and the obtained power balance \ref{eq:PowBalNI} with the appropriate boundary efforts and flows. However, since the coefficients of the operator depend explicitly on the effort variable $e_s=\rho T$, and not only on the energy variables $(\den, \vel,s)$, then, this is a pseudo Stokes-Dirac structure.\hfill\hfill\qed
\end{rmk}
\begin{rmk}
	Using some simple mathematical operation the term $\ten:\left[\frac{\ev}{\den}\n^{ \T}\right]$ in \eqref{eq:pf51} can be rewritten as $$\ten:\left[\frac{\ev}{\den}\n^{ \T}\right]=\frac{\ev}{\den}\cdot\left[\ten\cdot\n\right]$$
	Thus, for computational purposes the boundary  flows and efforts can be expressed as
	\begin{align}
	\fb=\begin{bmatrix}
	\left.-\left(\ev\cdot\n\right)\right|_{\Vb} \\ \left. -\vel\right|_{\Vb}\\ \left.-\left(\hf_s\cdot\n\right)\right|_{\Vb}
	\end{bmatrix} \text{ and } \eb=\begin{bmatrix}
	\left.\er\right|_{\Vb}\\\left[\ten\cdot\n\right]|_{\Vb}\\\left.T\right|_{\Vb}
	\end{bmatrix},
	\end{align}
	respectively.\hfill\hfill\qed
\end{rmk}

Now, consider $\mathbf{f}_d=\mathcal{G}_T\es$ and $\mathbf{e}_d=S_T\mathbf{f}_d$ as the flow and effort associated with the entropy diffusion by heat flux. Thus, an alternative representation for the model described in \eqref{eq:PHNonisen} is given by:
\begin{align}
\begin{bmatrix}
\dt{\den}\\\dt{\vel}\\\dt{\s}\\\mathbf{f}_d
\end{bmatrix}&=\begin{bmatrix}
0 & -\div & 0 &0\\
-\grad &-\frac{1}{\den}G_{\vor} & \Jt&0\\
0 & -\Jt^* & \mathcal{Q}_T & -\mathcal{G}^{*}_T\\
0 & 0 & \mathcal{G}_T &0
\end{bmatrix}\begin{bmatrix}
\er\\\ev\\\es\\\mathbf{e}_d
\end{bmatrix}
\end{align}
where the boundary flow and effort are given by
\begin{align}
\fb=\begin{bmatrix}
\left.-\left(\ev\cdot\n\right)\right|_{\Vb} \\ \left. -\vel\n^{ \T}\right|_{\Vb}\\ \left.-\left(\mathbf{e}_d\cdot\n\right)\right|_{\Vb}
\end{bmatrix} \text{ and }\eb=\begin{bmatrix}
\left.\er\right|_{\Vb}\\\ten|_{\Vb}\\\left.T\right|_{\Vb}
\end{bmatrix},
\end{align}
respectively, satisfying the relationship in \eqref{eq:PowBalNI}.

%

\section{Isentropic fluid}\label{sec:Isen}

	In this section we describe the port-Hamiltonian formulation for ideal isentropic fluids. The governing equations are reduced to the continuity and motion equations described in \eqref{eq:Con1} and \eqref{eq:Mot1}, respectively. Similarly, the Gibbs equation is reduced to
	\begin{align}
	du=-pd\left(\frac{1}{\den}\right)\label{eq:du_ise}
	\end{align}  
	
	In isentropic fluids the internal energy is a function that depends only on the density, as shown in \eqref{eq:du_ise}.
	Then, the total energy is described as
	\begin{align}
	\Ha&=\int_{\V}\frac{1}{2}\den\vel\cdot\vel+\den u\left(\den\right)
	\end{align}
	and the fluid efforts $\mathbf{e}=\begin{bmatrix}
	\er & \ev^{ \T}
	\end{bmatrix}^{ \T}$ are given by 
	\begin{align}
	\begin{bmatrix}
	\er\\	\ev
	\end{bmatrix}&=\begin{bmatrix}
	\delta_{\den}\Ha\\ \delta_{\vel}\Ha
	\end{bmatrix}=\begin{bmatrix}
	\frac{1}{2}\vel\cdot\vel+h\\
	\den\vel
	\end{bmatrix}\label{eq:efforts2}
	\end{align}
	Then, the fluid dynamics can be expressed as
	\begin{subequations}
		\begin{align}
		\dt{\den}&=-\div{\ev}\\
		\dt{\vel}&=-\grad{\er}-\frac{1}{\den}G_{\vor}\ev-\frac{1}{\den} \Div{\ten}\label{eq:Mot2}
		\end{align}\label{eq:fluid3}
	\end{subequations}

	The term $\frac{1}{\den}\Div{\ten}$ in \eqref{eq:Mot2} represents the friction effects over the velocity field, given the fluid viscosity. In previous section, the velocity field and the entropy of the fluid are interconnected through the heat generated by this friction, by means of the operators $\Jt$ and $\Jt^*$. In this case, given the isentropic assumption, we can interpret $\frac{1}{\den}\Div{\ten}$ as the dissipation associated with heat generation as a consequence of the viscosity friction of the fluid.
	According to \cite{Villegas2006}, in infinite-dimensional port-Hamiltonian systems the dissipative terms are expressed as $\mathcal{G}^* S \mathcal{G}\mathbf{e}$ where $\mathcal{G}^*$ is the adjoint operator of $\mathcal{G}$, and $S=S^{ \T}\geq0$. Then, for isentropic fluids $\frac{1}{\den}\Div{\ten}$ can be expressed as a port-Hamiltonian dissipation term, as shown in the following Lemma.	.
	
	\begin{lem}\label{lem:3}
		Let be a viscous Newtonian fluid. Defining the operators $\Gw=\curl{\frac{\cdot}{\den}}$ and $\Gr=\div{\frac{\cdot}{\den}}$ and the corresponding adjoints  $\Gw^*=\frac{1}{\den}\curl{}$ and $\Gr^*=-\frac{1}{\den}\grad{}$. Then, the rate of velocity addition associated with the viscous tensor, $\frac{1}{\den}\Div{\ten}$, can be expressed as a dissipative port-Hamiltonian terms associated with the velocity effort, namely, 
		\begin{align}
		\frac{1}{\den} \Div{\ten}&=\Gt^* S_{\boldsymbol{\tau}} \Gt\ev
		\end{align}
		where $\Gt^*=\begin{bmatrix}
		\Gw^* & \Gr^*
		\end{bmatrix}$, $S_{\boldsymbol{\tau}}=\begin{bmatrix}
			\visc & 0 \\0 & \hat{\visc}
		\end{bmatrix}$  and $\Gt=\begin{bmatrix}
		\Gw \\ \Gr
		\end{bmatrix}$, with $\hat{\visc}=\frac{4}{3}\visc+\visd$.
	\end{lem}
	\begin{pf}
		The viscosity tensor of Newtonian fluids is described in \eqref{eq:Ntensor}. Then, applying the identities \eqref{eq:A4}-\eqref{eq:A6} we obtain
		\begin{align}
			\frac{1}{\den} \Div{\ten}=&\frac{1}{\den} \Div{\left[-\visc\left(\Grad{\vel}+\left[\Grad{\vel}\right]^{ \T}\right)\right]} \nonumber\\
			&+ \frac{1}{\den} \Div{\left[\left(\frac{2}{3}\visc-\visd\right)\left(\div{\vel}\right)I\right]}\nonumber\\
			=&\frac{1}{\den}\curl{\left[\visc\curl{\vel}\right]}\nonumber\\
			&-\frac{1}{\den}\grad{\left(\left(\frac{4}{3}\visc+\visd\right)\div{\vel}\right)}\nonumber\\
			=&\frac{1}{\den}\curl{\left[\visc\curl{\frac{\ev}{\den}}\right]}\nonumber\\
			&-\frac{1}{\den}\grad{\left(\hat{\visc}\div{\frac{\ev}{\den}}\right)}\label{eq:pf4}
		\end{align}
		Given that the curl operator is self-adjoint and the adjoint of divergence is minus the gradient, then, it is easy to check that $\Gw^*=\frac{1}{\den}\mathbf{curl}$ is the formal adjoint of $\Gw=\curl{\frac{\cdot}{\den}}$ and $\Gr^*=-\frac{1}{\den}\grad{}$ is the adjoint of $\Gr=\div{\frac{\cdot}{\den}}$. Thus, equation \eqref{eq:pf4} can be expressed as the sum of 2 dissipative terms, namely
		\begin{align}
			\frac{1}{\den} \Div{\ten}&=\Gw^*\mu\Gw\ev+\Gr^*\hat{\visc}\Gr\ev\label{eq:2frictions}\\
				&=\underset{\Gt^*}{
						\underbrace{\begin{bmatrix}
									\Gw^* & \Gr^*
									\end{bmatrix}}} 
				\underset{S_{\boldsymbol{\tau}}}{
					\underbrace{\begin{bmatrix}
							\visc & 0 \\ 0 &\hat{\visc}
							\end{bmatrix}}}
				\underset{\Gt}{
					\underbrace{\begin{bmatrix}
						\Gw \\ \Gr
						\end{bmatrix}}}\ev
		\end{align}
		where $S_{\boldsymbol{\tau}}$ satisfies the positive condition $S_{\boldsymbol{\tau}}=S_{\boldsymbol{\tau}}^{ \T}\geq 0$.\hfill\hfill\qed
	\end{pf}
	
	Note that $\Gt^* S_{\boldsymbol{\tau}} \Gt\ev$ can be expressed as the sum of two dissipations, as shown in \eqref{eq:2frictions}. The first dissipation, $\Gw^*\visc\Gw\ev$, describes the losses associated with the frictions generated by the fluid rotation or vorticity, and it is equal to 0  under an irrotational assumption. The second dissipation, $\Gr^*(\frac{4}{3}\visc+\visd)\Gr\ev$, describes the losses associated with the frictions generated by the dilatation or compression of the fluid, and it is equal to 0 under incompressible assumption. 
	
	\begin{proposition}\label{prop:isen}
		Let be an isentropic Newtonian fluid in a domain $\V$ with boundary $\Vb$. Considering the vorticity as a phenomena strictly intern, the governing equations can be expressed as the following port-Hamiltonian system with dissipation:
		\begin{align}
			\dt{\boldsymbol{x}}&=\left(\J-\mathcal{G}^*S\mathcal{G}\right)\mathbf{e}\label{eq:PHIsen}
		\end{align}
		where $\boldsymbol{x}=\begin{bmatrix}
		\den & \vel^{ \T}
		\end{bmatrix}^{ \T}$ is the state vector, $\mathbf{e}=\begin{bmatrix}
		\er & \ev^{ \T}
		\end{bmatrix}^{ \T}$are the fluid efforts, and
		\begin{align*}
		\small\J=\begin{bmatrix}
		0&-\div\\-\grad&-\frac{G_{\vor}}{\den}
		\end{bmatrix}, \mathcal{G}^*=\begin{bmatrix}
		0 & 0\\0 &\Gt^* 
		\end{bmatrix}, S=\begin{bmatrix}
		0 & 0\\0 &S_{\boldsymbol{\tau}} 
		\end{bmatrix},\mathcal{G}=\begin{bmatrix}
		0 & 0\\0 &\Gt
		\end{bmatrix}
		\end{align*}
		Satisfying the following relationship for the rate of change of the energy:
		\begin{align}
			\frac{d\Ha}{dt}\leq \int_{\Vb}\fb \cdot \eb
		\end{align}
		where $\fb=\left(\er-\ed\right)|_{\Vb}$ and $\eb=-\left(\ev\cdot\n\right)|_{\Vb}$ are the boundary flow and effort, respectively, with $\ed$ as the effort associated with the dissipation by dilatation and $\n$ the normal unitary outward vector to the boundary.
	\end{proposition}
	\begin{pf}
	Considering the Lemma \ref{lem:3}, the dynamics in \eqref{eq:fluid3} can be rewritten as
		\begin{subequations}
			\begin{align}
			\dt{\den}&=-\div{\ev}\\
			\dt{\vel}&=-\grad{\er}-\frac{1}{\den}G_{\vor}\ev-\Gt^*S_{\boldsymbol{\tau}}\Gt\ev
			\end{align}
		\end{subequations}
		Thus, regrouping terms the governing equations can be expressed as
		\begin{align*}
		\dt{\begin{bmatrix}
			\den\\\vel
			\end{bmatrix}}&=\left(\begin{bmatrix}
				0&-\div\\-\grad&-\frac{G_{\vor}}{\den}
	\end{bmatrix}-\begin{bmatrix}
	0 \\\Gt^* S_{\ten}\Gt
	\end{bmatrix}\right)\begin{bmatrix}
	\er\\\ev
	\end{bmatrix}
		\end{align*}
		Rewriting the term $\begin{bmatrix}
		0 \\ \Gt S_{\ten}\Gt
		\end{bmatrix}$ we obtain the port-Hamiltonian formulation described in \eqref{eq:PHIsen}.
		
		On the other hand, for the rate of change of the total energy we obtain
		\begin{align}
			\frac{d\Ha}{dt}=&\int_{\V}\mathbf{e}\cdot\dt{\boldsymbol{x}}=\int_{\V}\mathbf{e}\cdot\J\mathbf{e}-\mathbf{e}\cdot\mathcal{G}^{*}S\mathcal{G}\mathbf{e}\nonumber\\
			=&-\int_{\Vb}\er\left(\ev\cdot\n\right)-\int_{\V}\mathbf{e}\cdot\mathcal{G}^{*}S\mathcal{G}\mathbf{e}\label{eq:dtH}
		\end{align} 
	Defining $\mathbf{f}_{R}=\begin{bmatrix}
	\fw^{ \T} & \fd
	\end{bmatrix}^{ \T}$ and $\mathbf{e}_{R}=\begin{bmatrix}
	\ew^{ \T} & \ed
	\end{bmatrix}^{ \T}$ as the flows and efforts associated with the dissipations, where $\fw=\Gw\ev$, $\fd=\Gr\ev$, $\ew=\visc\fw$ and $\ed=(\frac{4}{3}\visc+\visd)\fd$, we obtain $\mathbf{e}\cdot\mathcal{G}^{*}S\mathcal{G}\mathbf{e} =\ev\cdot\Gw^*\ew+\ev\cdot\Gr^*\ed$. Then, considering the vorticity equal to 0 in the boundaries, the equation \eqref{eq:dtH} can be rewritten as
	\begin{align}
	\frac{d\Ha}{dt}=&-\int_{\Vb}\er\left(\ev\cdot\n\right)-\int_{\V}\ev\cdot\Gw^*\ew+\ev\cdot\Gr^*\ed \nonumber\\
	=&-\int_{\Vb}\er\left(\ev\cdot\n\right)-\int_{\V}\Gw\ev\cdot\ew+\Gr\ev\cdot\ed \nonumber\\
	&+\int_{\Vb}\frac{\ed}{\den}\left(\ev\cdot\n\right)\nonumber\\
	=&-\inner{\Gt\mathbf{e}_R,\mathbf{e}_R}{\V}-\int_{\Vb}\left(\er-\frac{\ed}{\den}\right)\left(\ev\cdot\n\right)\nonumber\\
	=&-S_{\boldsymbol{\tau}}\inner{\mathbf{f}_R,\mathbf{f}_R}{\V}-\int_{\Vb}\left(\er-\frac{\ed}{\den}\right)\left(\ev\cdot\n\right)\label{eq:dtH2}
	\end{align}
	Given that $S_{\boldsymbol{\tau}}\geq0$, then, from \eqref{eq:dtH2} we obtain the inequality  $\frac{d}{dt}\Ha\leq\int_{\Vb}\fb\cdot\eb$, where $\fb=\left(\er-\dfrac{\ed}{\den}\right)|_{\Vb}$ and $\eb=-\left(\ev\cdot\n\right)|_{\Vb}$.\hfill\hfill\qed
\end{pf}
	
	Note that, considering different assumptions the model of fluid proposed in \eqref{eq:PHIsen}, corresponds  to port-Hamiltonian models of isentropic fluids described in previous works. For example, under an irrotational assumption operators $G_{\vor}$, $\Gw$ and $\Gw^{*}$ disappear, obtaining the fluid model described by \cite{Matignon2013}. On the other hand, for inviscid fluids, the operator $\mathcal{G}^{*}S\mathcal{G}$ is equal to 0, and then, the port-Hamiltonian system \eqref{eq:PHIsen} is equivalent to the model proposed in \cite{VanderSchaft2002}.  

\section{Two-dimensional fluids}\label{sec:2D}

The cross product and  the curl operator are three-dimensional mathematical operations. Thus, for two-dimensional fluids we need to properly define the terms associated with these operators.

Let us denote by $\{x_1,x_2\}$ the variables associated with the axes of a two-dimensional velocity field $\vel=\begin{bmatrix}
v_1 & v_2
\end{bmatrix}^{ \T}$. Then, the vorticity $w$ is a scalar 
defined as $\vorD=-\partial_{x_2}v_1 +\partial_{x_1}v_2$. For convenience we rewrite $w$ as
\begin{align}
w&=-\div{\left[W\vel\right]}
\end{align}
where $W=\begin{bmatrix}
0 & -1\\1& 0
\end{bmatrix}$ is a rotation matrix.

Then, the  {Gyroscope} in a two-dimensional velocity field is defined as \citep{Carodo-Ribeiro2016}:
\begin{align}
	G_{\vor}&=\vorD W=\begin{bmatrix}
	0 & -\vorD\\\vorD&0
	\end{bmatrix}\label{eq:Gyro2D}
\end{align}

On the other hand, with respect to the dissipative terms of the viscosity tensor,  operators $\Gw$ and $\Gw^{*}$ for two-dimensional fluids are defined as:
\begin{align}
\Gw& {=\begin{bmatrix}
	-\partial_{x_2} & \partial_{x_1}
	\end{bmatrix}\frac{\cdot}{\den}}=-\div{\left[W\frac{\cdot}{\den}\right]}\label{eq:Gw2D}\\
\Gw^{*}& {=\frac{1}{\den}\begin{bmatrix}
	\partial_{x_2} \\ -\partial_{x_1}
	\end{bmatrix}}=\frac{1}{\den}W^{ \T}\grad{}\label{eq:Gwa2D}
\end{align}

Thus, given the operator definitions in \eqref{eq:Gyro2D}-\eqref{eq:Gwa2D},  the port-Hamiltonian formulations in Propositions \ref{prop:nonisen} and \ref{prop:isen} can be used to describe non-isentropic and isentropic two-dimensional fluids, respectively.

 {In the case of 1D fluids, all terms associated with the vorticity disappear, and $\div=\grad=\partial_{x_1}$. Thus, the fluid model \eqref{eq:PHNonisen} is equivalent to the model described in \cite{Altmann2017}, neglecting the reactive part. Similarly, the model described in Proposition \ref{prop:isen} correspond to the models used in \cite{Kotyczka2013} and \cite{Macchelli2017}.}

\section{Conclusion}\label{sec:Conclu}

A pseudo-PH  formulation for 3D non-isentropic Newtonian fluids was presented for non-reactive flows. Similarly, under an isentropic assumption, the transformation of kinetic energy into heat by viscosity friction is described as dissipative terms associated with fluid rotation and compression, obtaining a dissipative-PH model for three-dimensional isentropic fluids. These models present a general formulation for non-reactive compressible flows, i.e., a description for inviscid or irrotational fluids can be derived from the proposed models under the corresponding assumptions in the PH structure.
Moreover, we have described the necessary considerations on the operators  used in the proposed models for the case of two-dimensional and one-dimensional fluids, obtaining formulations for fluids models equivalent to those found in the literature.


\bibliography{FluidReferences}             
                                                   







\appendix
\section{Nomenclature  and Useful identities}    
The nomenclature used in this paper is summarized in the next Table.
\begin{table}[h]
	\caption{Nomenclature}
	\centering
	\begin{tabular}{cl}
		Symbol & Description \\
		\hline
		$ \T$& Transpose\\
		$\vel\cdot\un$ & Scalar product between 2 vectors, $\vel^{ \T}\un$.\\
		$\vel\times\un$ & Cross product\\
		$\ten:\tena$ & Scalar product between 2 tensors, $Tr\left(\ten^{ \T}\tena\right)$.\\
		$\div{\un}$ & Divergence of vector $\un$.\\
		$\grad{f}$ & Gradient of scalar $f$.\\
		$\curl{\un}$& Curl or rotational of $\un$.\\
		$\Grad{\un}$& Gradient of vector $\un$.\\
		$ \Div{\tena}$& Divergence of tensor $\tena$.\\
		$\left\|\vel\right\|^2_{X}$ & Square of the weighted Euclidean norm, $\vel^TX^T\vel$.\\
		$\int_{\V}f$ & Integral in domain $\V$, $\int_{\V}f d\V$\\
		$\int_{\Vb}f$ & Integral in boundary $\Vb$, $\int_{\Vb}f \Vb$\\
		\hline
	\end{tabular}
\end{table}

Additionally, the set of mathematical identities \cite[Appendix A]{Bird2015} used in this work are described below:
\begin{align}
&\un\cdot\Grad{\un}=\grad{\left(\frac{1}{2}\un\cdot\un\right)} +\left[\curl{\un}\right]\times\un\label{eq:A1}\\
&\tena:\Grad{\un}=\div{\left[\tena\cdot\un\right]}-\un\cdot \Div{\tena}\label{eq:A2}\\
&\div{\left[f\un\right]}=\left[\grad{f}\right]\cdot\un+f\div{\un}\label{eq:A3}\\
& \Div{\left[\Grad{\un}\right]}=\grad{\left(\div{\un}\right)}-\curl{\left[\curl{\un}\right]}\label{eq:A4}\\
& \Div{\left[\Grad{\un}^{ \T}\right]}=\grad{\left(\div{\un}\right)}\label{eq:A5}\\
& \Div{\left[\left(\div{\un}\right)I\right]}=\grad{\left(\div{\un}\right)}\label{eq:A6}
\end{align}
where $f$ is a scalar, $\un$ is a vector and $\tena$ is a symmetric second order tensor.

\section{Useful properties}

In this section we describe some useful properties used in this paper.

\begin{thm}[Gauss Divergence Theorem]
	Let be a close domain $\V$, enclosed by the boundary surface $\Vb$, then
	\begin{align}
	\int_{\V}\div \un=\int_{\Vb}\un\cdot\n \label{eq:B1}
	\end{align}
\end{thm}
\begin{pf}
	See \cite[p. 704]{Bird2015}.\hfill\hfill \qed
\end{pf}

\begin{thm}[Adjoint of $\div$]
Let be the Hilbert space of the square integrable scalar functions, denoted by $\mathscr{H}_0=L^2(\V,\mathbb{R})$, and the Hilbert space of the square integrable vector functions, denoted by $\mathscr{H}_1=L^2(\V,\mathbb{R}^n)$. Given the operators $\div : \mathscr{H}_1\to \mathscr{H}_0$ and $\grad : \mathscr{H}_0\to \mathscr{H}_1$, where $-\grad$ is the formal adjoint of $\div$, then,
	\begin{align}
	\int_{\V}f\div{\un}+\int_{\V}\grad{f}\cdot\un=\int_{\Vb}f\left(\un\cdot\n\right)\label{eq:B2}
	\end{align}
\end{thm}
\begin{pf}
Denote by $\inner{f_1,f_2}{\mathscr{H}_0}=\int_{\V}f_1f_2$ and $\inner{\un_1,\un_2}{\mathscr{H}_1}=\int_{\V}\un_1\cdot\un_2$ the inner products in $\mathscr{H}_0$ and $\mathscr{H}_1$, respectively. Then,
\begin{align}
\inner{f,\div\un}{\mathscr{H}_0}+\inner{\grad f,\un}{\mathscr{H}_1}&=\int_{\V}f\div{\un}+\int_{\V}\grad{f}\cdot\un\nonumber\\&=\int_{\V}f\div{\un}+\grad{f}\cdot\un\nonumber\\
&\overset{\text{\eqref{eq:A3}}}{=}\int_{\V}\div{f\un}\nonumber\\
&\overset{\text{\eqref{eq:B1}}}{=}\int_{\Vb}f\left(\un\cdot\n\right)\nonumber
\end{align}
where for boundary conditions equal to 0, the relationship $\inner{f,\div\un}{\mathscr{H}_2}=\inner{-\grad f,\un}{\mathscr{H}_1}$ is obtained.\hfill\hfill\qed
\end{pf}

\begin{thm}
	Let be the Hilbert space of the square integrable vector functions $\mathscr{H}_1$ and the Hilbert space of the square integrable symmetric second order tensor functions, denoted by $\mathscr{H}_2=L^2\left(\V,\mathbb{R}^{n\times n}\right)$. Given the operators $\Div : \mathscr{H}_2\to \mathscr{H}_1$ and $\Grad : \mathscr{H}_1\to \mathscr{H}_2$,
	then, for one symmetric tensor $\tena\in\mathscr H_2$ and one vector $\un\in\mathscr H_1$, we obtain the following relationship
	\begin{align}
	\int_{\V}\left[ \Div{\tena}\right]\cdot\un+\int_{\V}\tena:\Grad{\un}=\int_{\Vb}\left[\tena\cdot\un\right]\cdot\n\label{eq:B3}
	\end{align}
\end{thm}
\begin{pf}
	Denote by $\inner{\un_1,\un_2}{\mathscr{H}_1}=\int_{\V}\un_1\cdot\un_2$ and $\inner{\tena_1,\tena_2}{\mathscr{H}_2} $ $=\int_{\V}\tena_1 : \tena_2$ the inner products in $\mathscr{H}_1$ and $\mathscr{H}_2$, respectively. Then, for a vector $\un\in\mathscr H_1$ and a symmetric tensor $\tena\in\mathscr H_2$, we obtain
	\begin{align}
	\inner{\Div \tena,\un}{\mathscr{H}_1}+\:\:\:\:\:\:\:\:\:\:\:\:\:\:&\nonumber\\ \inner{\tena,\Grad\un}{\mathscr{H}_2}&=\int_{\V}\left[\Div\tena\right]\cdot\un+\int_{\V}\tena:\Grad\un\nonumber\\
	&=\int_{\V}\left[\Div\tena\right]\cdot\un+\tena:\Grad\un\nonumber\\
	&\overset{\text{\eqref{eq:A2}}}{=}\int_{\V}\div\left[\tena\cdot\un\right]\nonumber\\
	&\overset{\text{\eqref{eq:B1}}}{=}\int_{\Vb}\left[\tena\cdot\un\right]\cdot\n\nonumber
	\end{align}
	where for boundary conditions equal to 0, the relationship $\inner{\Div \tena,\un}{\mathscr{H}_1}=\inner{\tena,-\Grad\un}{\mathscr{H}_2}$ is obtained.\hfill\hfill\qed
\end{pf}

\end{document}